\documentclass[twocolumn,showpacs,amsmath,amssymb,nofootinbib]{revtex4}

\newcommand{\ds}{\displaystyle}

\newcommand{\be}{\begin{equation}}
\newcommand{\en}{\end{equation}}
\newcommand{\bea}{\begin{eqnarray}}
\newcommand{\ena}{\end{eqnarray}}

\usepackage{graphicx}
\usepackage{bm}

\begin{document}

\title{Extended closed inflationary universes}

\author{ Sergio del Campo\footnote{E-mail address: sdelcamp@ucv.cl}   }
\affiliation{ Instituto de F\'\i sica, Pontificia Universidad
Cat\'olica de Valpara\'\i so, Av. Brasil 2950, Casilla 4059,
Valpara\'\i so, Chile.}
\author{Ram\'on Herrera\footnote{E-mail address: rherrera@unab.cl}}
\affiliation{Departamento de Ciencias F\'\i sicas, Universidad
Andr\'es Bello, Av. Rep\'ublica 237, Santiago, Chile.}

\date{\today}

\begin{abstract}
In this paper we study a type of  model for closed inflationary
universe models using  the Jordan-Brans-Dicke theory. Herein we
determine and characterize the existence of $\Omega>$1, together
with the period of inflation. We have found that our model, which
takes into account a Jordan-Brans-Dicke type of theory, is less
restrictive  than the one used in Einstein's theory of general
relativity. Our results are compared to those found in Einstein's
theory of Relativity.
\end{abstract}

\maketitle

\section{\label{sec:level1} Introduction}

The existence of Doppler peaks and their respective localization
tend to confirm the inflationary model, which is associated with a
flat universe with $\Omega=1.02\pm\,0.02$, as corroborated by the
existence of an almost scale invariant power spectrum, with
$n_{s}=0.97\pm\,0.03$\cite{CMB}.

The recent temperature anisotropy power spectrum measured with the
Wikinson Microwave Anisotropy Probe (WMAP) at high multipoles is
in agreement with an inflationary $\Lambda$- dominated CDM
cosmological model. However, the low order multipoles have lower
amplitudes than expected from this cosmological
model\cite{Bernett}. Perhaps, these amplitudes may indicate the
need for new physics. Speculations for explaining this discrepancy
has been invoked in the sense than the low quadrupole observed in
the CMB is related to the curvature scale\cite{Estathi}.

Due to this it may be interesting to consider other inflationary
universe models where spatial curvature is taken into account. In
fact, it is interesting to check if the flatness, in the curvature
as well as in the spectrum  are indeed reliable and robust
predictions of inflation\cite{Lindeclosed}.

In the context of an open scenario, it is assumed that the
universe has a lower-than-critical matter density and, therefore,
a negative spatial curvature. Several
authors~\cite{re3,re4,re5,re6}, following previous speculative
ideas~\cite{re1,re2}, have proposed possible models, in which open
universes may be realized, and their consequences, such as density
perturbations, have been explored~\cite{UMRS}. The only available
semi-realistic model of open inflation with $1-\Omega\ll\,1$  is
rather unpleasant; since it requires a fined-tuned potential of
very peculiar shape~\cite{re6,delherr}. Very recently the
possibility to create an open universe from the perspective of the
brane-world scenarios has been considered~\cite{MBPGD}.

The possibility of having inflationary universe models with
$\Omega>$1 has  also been
considered\cite{Lindeclosed,White,Ellis}. Particularly, this case
has been marginally indicated by the WMAP recent
observations\cite{Estathi}.

In this paper we study inflationary closed universe models in
which a Jordan-Brans-Dicke (JBD) theory is taken into
account\cite{JBD}. Following the line described by
Linde\cite{Lindeclosed}, we considered two situations: in the
first, the inflaton potential $V(\sigma)$ is constant. More
precisely, in this case the model is based on a step-like
effective potential in which $V(\sigma)=$ 0 at $\sigma<$ 0 and
$V(\sigma)=V=const$ for the range of $\sigma$ given by
0$<\sigma<\sigma_o$, where $\sigma_{o}$ is its initial value.
Close to this value ($\sigma=\sigma_o$) we will assume that the
potential sharply rises to infinitely large values. In a universe
with $\Omega > 1$, and in order to make inflation short, one could
consider that the effective potentials at the Planck density are
extremely steep. In general, such a universe will not typically
enter the inflationary regime and will collapse quickly. This
suggests that most of the universes described by the models that
use extremely steep potentials  will not inflate at all, or will
expand by a factor much less than 60 e-folds. In this way, most of
the universe will be short living. We like to address this
particular problem here; the other effective potential that we are
considering is a chaotic potential of the form
$V(\sigma)=\lambda_n\sigma^n/n$. If we agree with the probability
of creation of an inflationary universe, it is suggested that it
is much more natural for the universe to be created with a density
very closed to the Planck density.

In the case of Einstein`s theory of general relativity with a
constant scalar inflaton potential $V(\sigma)$, the inflaton field
$\sigma$ stops before it reaches the value $\sigma=0$, whose value
is necessary for ending inflation, and hence inflation would
continue forever\cite{Lindeclosed}. Differently, in a JBD type of
theory, it is possible to solve this (graceful exit) problem. As
we will see, the JBD parameter $\omega$ and the initial condition
of the JBD field make it possible that the inflaton field reaches
the value $\sigma=0$ and thus solving the graceful exit problem.

The structure of the paper is  as follows: In Sec. II we present
the action and equations for the Jordan-Brans-Dicke theory. In
Sec. III we determine the characteristics of the closed
inflationary universe models that are produced with a potential
$V(\sigma)$= constant. In Sec. IV we determine closed inflationary
universe models for chaotic potentials of the form
$V(\sigma)=\lambda_n\sigma^n/n$. We give an estimation of the
scalar density perturbation $\delta\rho/\rho$. At a different
level, our results are compared with those obtained with
Einstein`s theory of gravity. Finally, conclusions and some
discussions are presented in Sec. V.


\section{\label{sec:level2}The cosmological equations in JBD theory }

We consider the effective action $S$ given by
 \be \hspace{0.cm} \ds
S\,=\,\int{d^{4}x\,\sqrt{-g}}\,\left
[\,\frac{1}{2}\,\varepsilon\,\phi^{2}\,R\,
-\,\frac{1}{2}\,\partial_{\mu}\phi \,\partial^{\mu}\phi
-{\cal{L}}(\sigma)\right], \label{ac1}
 \en
where
$${\cal{L}}(\sigma)\,=\,\frac{1}{2}\partial_{\mu}\sigma
\partial^{\mu}\sigma\,-\, V(\sigma),$$
and $R$ is the Ricci scalar curvature, $\phi$ is the JBD scalar
field, and $ \varepsilon$ is a dimensionless coupling constant
that, in terms of the JBD parameter
$\omega=\frac{1}{4\,\varepsilon}$. $V(\sigma)$is an effective
scalar potential associated with the inflaton field, $ \sigma$.

The Friedmann-Robertson-Walker metric is described by
 \be \ds
d{s}^{2}\,=\, d{t}^{2}\,-\, a(t)^{2}\, \,\, d\Omega^{2}_{k}\,\,,
\label{met} \en where $a(t)$ is the scale factor, $t$ represents
the cosmic time and $d\Omega^{2}_{k}$ is the spatial line element
corresponding to the hypersurfaces of homogeneity, which could
represent a three-sphere, a three-plane or a three-hyperboloid,
with values $k$=1, 0, -1, respectively. From now on  we will
restrict ourselves to the case $k$ = 1 only.

Using the metric~(\ref{met}), and $k$=1, in the
action~(\ref{ac1}), we obtain the following field equations: \be
\ds
\ddot{\sigma}\,=\,-3\,\frac{\dot{a}}{a}\,\dot{\sigma}\,-\,\frac{dV}{d\sigma}\,\,\label{ec1},
\en \be \ds
\ddot{\phi}=-3\frac{\dot{a}}{a}\dot{\phi}-\frac{\dot{\phi}^{2}}{\phi}
-\frac{1}{1+6\varepsilon}\left[\frac{\dot{\sigma}^{2}}{\phi}-\frac{4}{\phi}V(\sigma)
\right]\,\, \label{ec2},
 \en
\be
\frac{\dot{a}^2}{a^2}+2\frac{\dot{a}\dot{\phi}}{a\phi}=-\frac{1}{a^2}+
\frac{1}{3\varepsilon\phi^2}\left[\frac{\dot{\phi}^2}{2}+\frac{\dot{\sigma}^2}{2}+V(\sigma)
\right],
\en

and \be \ds
\ddot{a}=2\frac{\dot{a}\dot{\phi}}{\phi}-\frac{a}{3\varepsilon\phi^{2}}\left[
\dot{\phi}^{2}+\frac{1+3\varepsilon}{1+6\varepsilon}\dot{\sigma}^{2}-
\frac{1-6\varepsilon}{1+6\varepsilon}V(\sigma)\right] ,\label{ec3}
\en

where the dots over $\phi$ and $a$ denotes derivatives with
respect to the time $t$. For convenience  we will use from now on
units where c = $\hbar$=M$_{p}$=G$^{-1/2}$ = 1. Note that this set
of equation reduces itself to the set of Einstein`s field Eqs., in
the limits $\varepsilon\longrightarrow$ 0 and
$\varepsilon\phi^2\longrightarrow\,M_p^2/8\pi=1/8\pi$.

\section{\label{sec:level2}Closed inflationary universe model with $V=Cte.$ }

Following Linde\cite{Lindeclosed},  let us consider  for
simplicity a toy model with the following step-like effective
potential: $V(\sigma)=$ 0 at $\sigma<$ 0; $V(\sigma)=V=const$ at
0$<\sigma<\sigma_o$ (here $\sigma_{o}$ is the initial value of the
inflaton field). We shall also assume that the effective potential
sharply rises to infinitely large value in a small vicinity of
$\sigma=\sigma_o$. As Linde mentions in Ref.\cite{Lindeclosed}
this model is very well suited for the description of a certain
version of an F-term hybrid inflation in supergravity. In such
models the effective potential is nearly constant at small
$\sigma$ and it exponentially rises at large $\sigma$. We should
mention that the physical background of this model is inspired by
Linde's Hybrid inflation model\cite{LindeHy}. Therein, the
inflationary plateau is due to the vacuum density associated with
some other so-called "waterfall" field which is coupled to the
inflaton field. Here, the plateau is slightly tilted due to
radiative corrections so the field will continue to roll down, and
in addition  the end of inflation occurs at some critical value
$\sigma_{c}>0$, differing from our case in which inflation ends at
$\sigma = 0$. But, fundamentally, the models are rather similar.
In fact, for large values of $\sigma$, supergravity corrections
introduce a quartic term which grows rapidly and resembles very
much the abrupt cliff mentioned above.

Suppose now that a close universe described by these models
appeared "from nothing" in a state with the field
$\sigma\geq\sigma_o$
 at the point with
$\dot{a}=0$, $\dot{\sigma}=0$,
$\dot{\phi}=\sqrt{\beta_o}\dot{\widetilde{\phi_o}}$ and the
potential energy density equal to $V_{o}$ (with  $ V_{o}\geq\,V$).
Here, the time derivative of the JBD scalar $\dot{\phi}$ has an
initial condition that depends on $\beta_{o}$  and
$\dot{\widetilde\phi_{o}}$.  The  value of $\beta_o$ permits us to
establish when  inflation may occur, by means of  the probability
of creation of an inflationary universe model. This restricts the
initial value of the inflaton field, $\sigma_o$, and also, the
value of $\dot{\widetilde\phi_o}$ which represents a fraction of
the initial value of $\dot{\phi}$. We choose this particular
initial value for $\dot{\phi}$ in order that, in each one of the
successive intervals of the inflaton field, $\Delta\sigma$, and in
each one of the  successive intervals of the inflationary process,
these quantities do not depend on $\beta_o$.

It will be convenient to represent  $V_o$ as
$$
V_o=V ^{\ast}-\Delta\,V ^{\ast},
$$
where $V ^{\ast}=Cte.=\frac{3V}{2(1+3\varepsilon)}$. Note that
this quantity represents an upper bound on $V_o$ which occurs when
$\Delta\,V ^{\ast} \rightarrow$ 0. The definition of $V ^{\ast}$
in term of $V$ and $\varepsilon$ allows us to study the energy
density in the vicinity of $\sigma=\sigma_o$ and the one
established from energy conservation\footnote{In the case of GR theory:\\
The energy conservation
$V_o+0=\frac{\dot{\sigma}^2}{2}+V\Longrightarrow\dot{\sigma}^2=2(V_o-V)$
in Eq. for $\ddot{a}$ i.e.
$$
\ddot{a}=\frac{8\pi}{3}\,a(V-\dot{\sigma}^2)=\frac{8\pi}{3}\,a(3V-2V_o)
$$
Static universe $\ddot{a}=0\Longrightarrow\,V_o=3V/2$; Colapses
$\ddot{a}<0\Longrightarrow\,V_o>3V/2$; Inflation
$\ddot{a}>0\Longrightarrow\,V_o<3V/2$. In general
$V_o=3V/2-\Delta\,V$ with $\Delta\,V=0,<0,>0.$
\\
\\
In the case of JBD Theory (analogy):\\
The energy conservation
$V_o+0=\frac{\dot{\sigma}^2}{2}+V\Longrightarrow\dot{\sigma}^2=2(V_o-V)$
in Eq.(6), we find
$$
V_o=\frac{3V}{2(1+3\varepsilon)}-\Delta\,V^{\ast}=V^{\ast}-\Delta\,V^{\ast}.
$$} and the
equation(\ref{ec3}). Note that this value of $V^{\ast}$ yields the
appropriate Einstein's General Relativity limit for
$\varepsilon\longrightarrow\,0$. Here, $\Delta\,V ^{\ast}$ is a
small quantity that becomes a function of the quantities
$\varepsilon$, V and $\beta_o$.

Then, the field $\sigma$ instantly would fall down to the plateau
($V(\sigma)=V$), and its potential energy density would become
converted into the kinetic energy,
$\frac{\dot{\sigma}^2}{2}=\frac{(1-6\varepsilon)V}{2(1+3\varepsilon)}-\Delta\,V^\ast$,
and thus the velocity of the field $\sigma$ would be given with
$\dot{\sigma}=-\sqrt{(1-6\varepsilon)V/(1+3\varepsilon)-2\Delta\,V^\ast}$.
Since this happens nearly instantly, one still has $\dot{a}=0$,
$\dot{\phi}=\sqrt{\beta_o}\dot{\widetilde{\phi_o}}$ and
$\dot{\sigma}=-\sqrt{(1-6\varepsilon)V/(1+3\varepsilon)-2\Delta\,V^\ast}$
at these times. These values could be considered  initial
conditions when solving the set of Eqs.(\ref{ec1})-(\ref{ec3}),
for $V(\sigma)$=V=const in the interval $0<\sigma<\sigma_o$. Since
the field $\sigma$ instantly falls down to the plateau (V=const),
its potential energy density is converted into kinetic energy.
This causes  $\dot{\sigma}$ to change from a zero value to a
negative constant value.  During this change, we see from the
Friedmann equation that $\dot{a}=0$ and $\phi=const.=\phi_o$.
After this period,  the inflaton field $\sigma$ instantly falls
down to the plateau and  the evolution of the JBD field $\phi$
monotonically increases to some constant value, which we denote by
$\phi_{N}$. We fixe  this value to the actual value of the Planck
mass times $\sqrt{\omega/2\pi}$  (recall that
$\phi_N^2=M_p^2/8\pi\varepsilon\,$ which becomes
$\phi_N^2=\omega/2\pi=const.$ in unit with $M_p=1$).
 In Fig. 1 we show the temporal evolution of the JBD scalar field.

\begin{figure}[ht]
\includegraphics[width=3.0in,angle=0,clip=true]{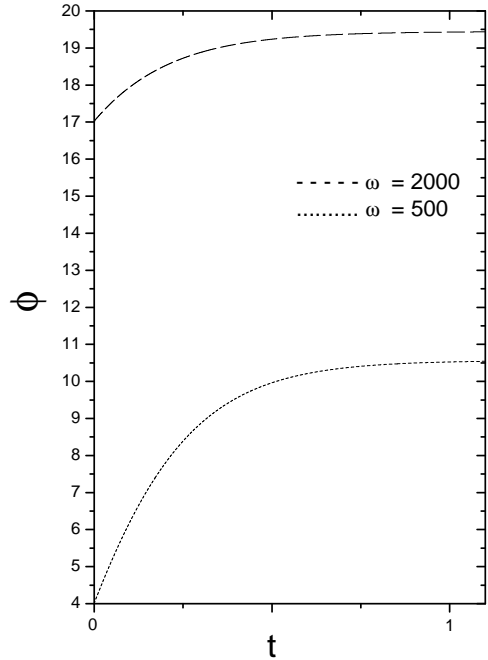}
\caption{For the model $V= const$ we have plotted the JBD field
$\phi$(t) as a function of the time $t$, for two different values
of the JBD parameter,  $\omega=500$ and $\omega=2000$.  We have
assumed the constant $V$ to be equal to one.} \label{fig33}
\end{figure}

In order  to clarify the conditions under which a closed universe
could remain static, or could collapse, or even could enter  an
inflationary regime, it is necessary to rewrite Eq.(\ref{ec3}) in
terms of $\Delta\,V^{\ast}$
$$ \ddot{a}=a\eta\left[2H-\frac{1}{3\varepsilon}\eta
\right]+\frac{2a(1+3\varepsilon)}{3\varepsilon\phi^2(1+6\varepsilon)}\Delta\,V^{\ast}\,,
$$
where $H=\dot{a}/a$ defines the Hubble expansion rate and
$\eta=\dot{\phi}/\phi=|\frac{\dot{G}}{2G}|\ll\,1$ where
$G=1/8\pi\varepsilon\phi^2$ is the gravitational ``constant". The
present value of $\eta$ is of the order of $\eta_{o}\leq
\,10^{-10}/yr$ which in Planck mass unit becomes
$\sim\,10^{-45}M_p$. In general, during inflation we have
$H\sim\,\sqrt{V}/M_{p}$ with $\sqrt{V}\sim\,10^{-5}M^2_p$ and we
assume that during inflation $H>\eta/6\varepsilon$.   This
assumption is justified, since   during inflation
$H_{inf}\sim\,10^{-5}M_p$. In our models, we take
$\eta_{inf}\sim\,10^{-11}M_p$, when we resolve the field equations
for the different values of the JBD parameter $\omega$, initial
values of the JBD field and the  time derivatives of the JBD
scalar $\dot{\phi}$. This approximation coincides with the
approach  used by the author in Ref.\cite{Lindeletter}, when  he
studies inflation in a JBD theory. In this form, the above
equation takes the form $\ddot{a}\simeq\,2a\eta\,H
+\frac{2a(1+3\varepsilon)}{3\varepsilon\phi^2(1+6\varepsilon)}\Delta\,V^{\ast}$.

In the particular case of $\Delta\,V^{\ast}=0$, different
possibilities exist as opposed  to what happens in Einstein`s GR
theory. In the case of $\eta\,H>0$, $\ddot{a}>0$ and
$3\dot{a}/a>0$, the value of $\dot{\sigma}$ decreases, and the
universe enters an inflationary phase. In the case of $\eta\,H<0$
, $\ddot{a}<0$ and $3\dot{a}/a<0$, the universe collapses. If
$H\eta+\frac{(1+3\varepsilon)}{3\varepsilon\phi^2(1+6\varepsilon)}\Delta\,V^{\ast}=0$,
the acceleration of the scale factor $\ddot{a}=0$ and the universe
remains static, as occurs in Einstein`s GR theory.

For
$H\eta+\frac{(1+3\varepsilon)}{3\varepsilon\phi^2(1+6\varepsilon)}\Delta\,V^{\ast}<0$,
the term $3\dot{a}/a<0$, which makes the motion of $\sigma$ even
faster with the consequence of a rapidly collapsing universe.

In the case of
$H\eta+\frac{(1+3\varepsilon)}{3\varepsilon\phi^2(1+6\varepsilon)}\Delta\,V^{\ast}>0$,
which means  $\ddot{a}>0$, we get $3\dot{a}/a\,>0$. The field
$\sigma$ rapidly decreases, and thus an inflationary phase is
generated.

Note also that, in the regime where $V$=const, the scalar field
$\sigma$ satisfies equation \be
\ddot{\sigma}\,=\,-3\,\frac{\dot{a}}{a}\,\dot{\sigma} ,\en which
implies that \be
\dot{\sigma}(t)=\dot{\sigma_o}\left(\frac{a_o}{a(t)} \right)^3.
 \label{campoesc}\en
Here $\dot{\sigma_o}$ is the initial velocity of the field
$\sigma$, immediately after it rolls down to the flat part of the
potential. But once expansion of the universe begins, this
expansion follows a simple power-law ( as opposed to the GR case
where $a\sim\,e^{H\,t}$, with $H=\sqrt{8\pi\,V/3}$, see
Ref.\cite{Lindeclosed}). The effect of the JBD scalar field in
this model is reflected in the change of the slope of the scalar
field $\sigma$, even if its form remains the same, as shown in
Figs. 2 and  3.  We could describe a power-law regime, (which
occurs if we  neglect $\dot{\phi}/\phi$ as compared with $H$,
$\ddot{\phi}$ as compared with $3H\dot{\phi}$ and
$\dot{\phi}^{2}$, $\dot{\sigma}^{2}$ as compared with the
effective potential $V(\sigma)$, and the term $1/a^2$ is rapidly
diluted ) and possible solutions would be given by
\cite{Lindeletter} \be
\frac{a(t)}{a_o}=\left[\frac{\phi(t)}{\phi_o}
\right]^{\gamma}=(\alpha\,t+1)^{\gamma}\label{eca}\,,
 \en
where
 $$\gamma=\frac{(1+6\varepsilon)}{4\varepsilon}\,\,;\,\alpha=\frac
 {4}{(1+6\varepsilon)\phi_{o}^2}\sqrt{\frac{\varepsilon\,\phi_{o}^2\,V}{3}},
 $$
in which $\phi_o$ is the initial value of the JBD.

 The resulting equation for the scale factor reads as follows:
 \be \ds
\ddot{a}\,=\,2\frac{\dot{a}\,\dot{\phi}}{\phi}-\frac{a\,\dot{\phi}^{2}}{3\varepsilon\phi^{2}}
+\frac{2\,a\,V\,\beta(t)}{3\varepsilon\phi^{2}}. \label{ecbeta}
\en

At this point, we will introduce a small time-dependent function
defined by
 \be
\beta(t)=\frac{1}{2V(1+6\varepsilon)}[(1-6\varepsilon)V-(1+3\varepsilon)\dot{\sigma}^2]\ll
1.
 \en

 Here, we would like to make a simple analysis of the solutions to Eq.(\ref{ecbeta}) for
 $\beta(0)\equiv\beta_o\ll$ 1, in which
$$
\beta_o=\frac{1}{2V(1+6\varepsilon)}[(1-6\varepsilon)V-(1+3\varepsilon)\dot{\sigma_o}^2]=
$$
\be=\,
\frac{(1+3\varepsilon)\Delta\,V^\ast}{(\,1\,+\,6\,\varepsilon)\,V}
\ll 1,\label{betacero}
 \en
when $a(t)$ increases  enough and an inflationary regime settles
in, the inflaton scalar field $\sigma$ gradually stops moving.
From Eq. (\ref{campoesc}) together with Eq. (\ref{eca}) we write
$\dot{\sigma}=\dot{\sigma}_o(\alpha t+1)^{-3\gamma}$, and $\sigma$
moves a finite distance, which in our case becomes
 \be
 \hspace{-0.1cm} \Delta\sigma_{\inf}=
  \frac{\dot{\sigma}_o}{\alpha(3\gamma-1)}\approx-
  \frac{(1+6\varepsilon)}{(3+14\varepsilon)}
  \sqrt{\frac{(1-6\varepsilon)3\,\varepsilon\,\phi_o^2}{(1+3\varepsilon)}}.
 \en
When $\varepsilon\longrightarrow$ 0 and
$\varepsilon\phi_o^2=Cte.=1/8\pi$, we obtain
$\Delta\sigma_{\inf}\approx-1/(2\sqrt{6\pi})$, which coincides
with the result  obtained in Einstein`s theory of Relativity
\cite{Lindeclosed}.

At the beginning of the process, when $a\approx\,a_o$,
$\phi\approx\phi_o$ and $\beta(t)\approx\beta_o$, Eq.
(\ref{ecbeta}) becomes
 \be \ds
\ddot{a}\,=\,2\frac{\dot{a}\,\dot{\phi}}{\phi_o}-
\frac{a_o\,\dot{\phi}^{2}}{3\varepsilon\phi_o^{2}}
+\frac{2\,a_o\,V\,\beta_o}{3\varepsilon\phi_o^{2}} \label{ecbetao}
,\en and for small $t$ the solution of Eq. (\ref{ecbetao}), i.e.
the scale factor $a(t)$, is found to be given by \be a(t)\simeq
a_o\left(1+\frac{\beta_o}{3\,\varepsilon\,\phi_o^2}
\left[V-\frac{\dot{\widetilde{\phi_o}}^2}{2}\right]\,t^2\,+\,O(t^3)
\right).\en From Eq. (\ref{campoesc}) we find that at the time
interval $\Delta\,t_1$
 given by
   \be
\Delta\,t_{1}\approx\sqrt{\frac{2(1+6\varepsilon)\varepsilon\phi_o^2}
{(1-6\varepsilon)(2V-\dot{\widetilde{\phi_o}}^2)}}\,\,,
 \en
 the parameter $\beta$ becomes twice   $\beta_o$, when consequently the
 inflaton field $\sigma$ decreased by the amount
 \be
\Delta\sigma_{1}\sim\,\dot{\sigma}(0)\Delta\,t_{1}\approx-
\sqrt{\frac{2V(1+6\varepsilon)\varepsilon\phi_o^2}{(1+3\varepsilon)
(2V-\dot{\widetilde{\phi_o}}^2)}}\,\,.
 \en
This process continues, after a time
$\Delta\,t_2\approx\Delta\,t_1$, where now the field $\sigma$
decreases by the amount $\Delta\sigma_2\approx\Delta\sigma_1$,
implying that the rate of growth of the scale factor, $a(t)$
doubles. This process finishes when
$\beta(t)$=$(1-6\varepsilon)/2(1+6\varepsilon)$. Here, the
beginning of inflation is determined by the initial value of the
inflaton field given by
$$ \sigma_{\inf}\approx\sigma_o+
\sqrt{\frac{3(1-6\varepsilon)\varepsilon\phi_o^2}{(1+3\varepsilon)}}
\,\,\frac{(1+6\varepsilon)}{(3+14\varepsilon)}\,+
$$
\be
 +\sqrt{\frac{2V(1+6\varepsilon)\varepsilon\phi_o^2}{(1+3\varepsilon)
(2V-\dot{\widetilde{\phi_o}}^2)}}
\ln\left[\frac{(1+6\varepsilon)\beta_o}{(1-6\varepsilon)}\right].
\en
 Note that this expression indicates that our results are very sensitive to the
choice of  particular values of $\omega$, and $\phi_o$, apart from
the other parameters that enter  the theory. Note that in the
limit in which $\varepsilon\phi_o^2\longrightarrow\,1/8\pi$ and
$\varepsilon\,\longrightarrow\,0$, the above expression reduces
itself to $\sigma_{\inf}\approx\sigma_o+0.1+0.15\ln\beta_o$, where
now, in this limit, $\beta_o$ becomes
$\beta_o\longrightarrow(1-\dot{\sigma}_o^2/V)/2$. Since inflation
occurs in the interval $\sigma_{\inf}> 0$ and $\sigma=0$, we
should obtain for the initial value of the inflaton field
$$ \hspace{-2.5cm}\sigma_o>-
\frac{(1+6\varepsilon)}{(3+14\varepsilon)}
\sqrt{\frac{3(1-6\varepsilon)\varepsilon\phi_o^2}{(1+3\varepsilon)}}\,-
$$
\be
\hspace{1.cm}-\sqrt{\frac{2V(1+6\varepsilon)\varepsilon\phi_o^2}{(1+3\varepsilon)
(2V-\dot{\widetilde{\phi_o}}^2)}}
\ln\left[\frac{(1+6\varepsilon)\beta_o}{(1-6\varepsilon)}\right].
\en

We continue describing a model of quantum creation for a closed
inflationary universe model. According to Ref.\cite{LDC}, the
probability of the creation of a closed universe filled with a
scalar field $\sigma$, with an effective potential $V(\sigma)$ in
the theory of JBD for the case $\varepsilon\phi^2\gg\,V^{1/2}$ is
given by

\be
 P\,\sim\,e^{-2|S|}\,=\exp\left(\frac{-3\,(8\pi\varepsilon\phi^2)^2}{8\,V(\sigma)}
 \right).
 \en

We first estimate the conditional probability that the universe is
created with an energy density equal to
$V^\ast-(1+6\varepsilon)\beta_o\,V/(1+3\varepsilon)$, under the
condition that its energy density $V$ was smaller than $V^\ast$;
for the initial value of the JBD field $\phi=\phi_o$ we find
$$
 P\sim\,\exp\left(\frac{-3\,(8\pi\varepsilon\phi_o^2)^2}{8}
 \left[\frac{1}{V^\ast-\frac{(1+6\varepsilon)\beta_o\,V}{(1+3\varepsilon)}}-\frac{1}{V^\ast}    \right]\right)
$$
\be
 P\,\sim\,\exp\left(-\frac{(8\pi\varepsilon\phi_o^2)^2
 (1+6\varepsilon)^2\beta_o}{6(1+3\varepsilon)\,V} \right),
  \en
which implies that the process of quantum creation of an
inflationary universe model is  not exponentially suppressed by
 $\beta_o<\frac{6(1+3\varepsilon)\,V}{(1+6\varepsilon)^2(8\pi\varepsilon\phi_o^2)^2}$,
which in turn means that the initial value of the inflaton field
$\sigma$ must be bounded from below

 $$
\hspace{-3.0cm} \sigma_o
>-\sqrt{\frac{3(1-6\varepsilon)\varepsilon\phi_o^2}{(1+3\varepsilon)}}\,
 \,\frac{(1+6\varepsilon)}{(3+14\varepsilon)}\,-
$$
\be
\sqrt{\frac{2V(1+6\varepsilon)\varepsilon\phi_o^2}{(1+3\varepsilon)
(2V-\dot{\widetilde{\phi_o}}^2)}}
\ln\left[\frac{6(1+3\varepsilon)\,V}{(1-36\varepsilon^2)(8\pi\varepsilon\phi_o^2)^2}\right].
\en

Let us consider two different cases for the values
$\varepsilon\phi{_o}^2$ and $\omega$.

First, suppose  $V\sim\,10^{-11}$, $8\pi\varepsilon\phi_o^2=1/2$,
$\dot{\widetilde{\phi_o}}^2=10^{-12}$ and $\omega=500$. In this
case the probability of creation of an inflationary universe is
not suppressed by $\beta_o<2.4\cdot10^{-10}$. The value $\beta_o$
allows us to fix the initial value of the inflaton field. In this
case we have found that inflation begins when the field
$\sigma_{o}$ takes the value \be
 \sigma_o\,>\,-0.08-0.14\ln(2\cdot10^{-10})\sim\,3.0\,M_p\,.
 \en

After the initial inflation, the field $\sigma$ stops moving when
it passes the distance $|\Delta\sigma_{\inf}|\simeq$  0.08.
However, this result is a particular value that depends on the
value we assign to the parameter $\omega$ and the initial values
of $\phi_o$ and $\dot{\phi_o}$ for a given value of the effective
potential $V$.

Now, suppose that $V\sim\,10^{-11}$,
$8\pi\varepsilon\phi_o^2=1/8$,
$\dot{\widetilde{\phi_o}}^2=10^{-12}$ and $\omega=1000$. In this
case, the probability of creation of an inflationary universe is
not suppressed by $\beta_o<3.8\cdot10^{-9}$. And hence we found
that inflation begins when the initial value of the inflaton field
$\sigma$ takes the value \be
 \sigma_o\,>\,-0.08-0.15\ln(4\cdot10^{-9})\sim\,2.8\,M_p\,.
 \en
Note that in this case, the field $\sigma$ moves a finite distance
which becomes $|\Delta\sigma_{\inf}|\,\simeq$ 0.08.
 If this field stops before it reaches $\sigma$=0, then the
 universe inflates eternally. The same problem is found in
 Einstein's General Relativity model ($\Delta\sigma_{\inf}\simeq-1/2\sqrt{6\pi}=const.$).
  However, in
the context of the JBD theory,  the value  of
$\Delta\sigma_{\inf}$, is quite sensitive to the values we assign
to $\omega$, $\phi_o$ and $\dot{\widetilde{\phi_o}}$. Therefore,
as we will see, the problem that the universe inflated  for ever
disappears and the inflaton field reaches the value $\sigma$ =0,
for $\omega<2000$ and for some appropriate initial conditions of
the JBD field $\phi_o$ and $\dot{\widetilde{\phi_o}}$.  It has
been claimed that the values of $\omega$ should be greater than
2000, in order that the JBD theory  be consistent with the
astronomical observations. In the more general scalar-tensor
theory, $\omega$ can vary, depending on the scalar field $\phi$.
In cosmological models, based on  such theories, it has been
pointed out that there is generally an attractor mechanism that
drives $\omega$ to $\infty$ in the late cosmological
epochs\cite{Damour}. But, at an early time in the evolution of the
universe the value of $\omega(\phi)$ can be significantly
different. Here, information on the different cosmological epochs
may constrain the value of $\omega$ to the range
$10<\omega<10^{7}$ (see Ref.\cite{2003}). Therefore, in the early
universe it is possible to consider $\omega$ which is smaller than
the one used at the present epoch\cite{1999}.

The numerical solutions to the inflaton field $\sigma(t)$ are
shown in Fig.2, for  different values of the $\omega$ parameter.
Note that the interval of $\sigma_o$ at $\sigma_{\inf}$ increases
when the parameter $\omega$ decreases, but its shapes remain
practically unchanged. We should note here that, as long as we
decrease the value of the $\omega$ parameter, the quantity
$\sigma_o-\sigma_{\inf}$ increases,  and thus it permits
$|\Delta\sigma_{\inf}|$  to reach $\sigma=0$, even when they
themselves are not oscillations of the inflaton field in neither
of the two  theories presented. If the field $\sigma$ starts its
motion with sufficiently small velocity, inflation begins
immediately, in analogy with Einstein's GR theory. If it starts
with large initial velocity $\dot{\sigma_o}$, because it falls
from the value of $V(\sigma)=V^{\ast}-\Delta\,V^{\ast}$, the
universe does not present the inflationary period at any stage.

\begin{figure}[ht]
\includegraphics[width=3.0in,angle=0,clip=true]{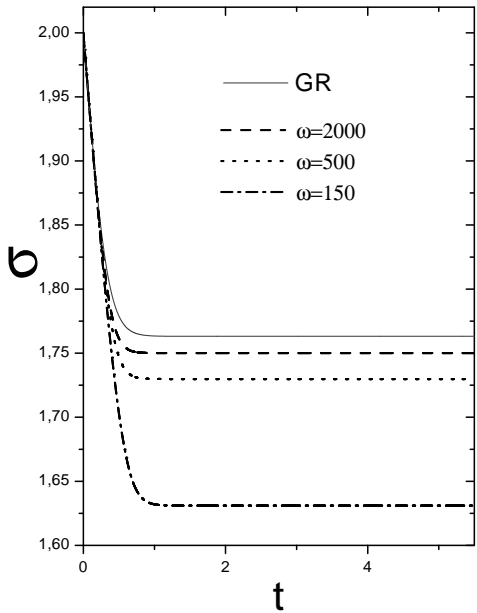}
\caption{For the model $V= const$ and the same value of
$\dot{\sigma_o}$, we have plotted the inflaton field $\sigma$(t)
as a function of the time, for three different values of the JBD
parameter, $\omega=150$, $\omega=500$ and $\omega=2000$,
respectively. Here we have assumed  the initial value for the JBD
field to be equal to $8\pi\varepsilon\phi_o^2=1/2$. GR represents
the same plot, however using Einstein's theory of Relativity. We
have assumed the constant $V$ to be equal one.} \label{fig1}
\end{figure}

\section{\label{sec:level2}Chaotic Inflation with $V=\lambda_n\,\sigma^n/n$ }

 Let us now consider chaotic inflation universe models as a next step.
In this model the main idea was to study all possible initial
conditions in the universe, including those where the scalar field
outside of the minimum of the scalar field potential\cite{LI}. In
this sense, in order to study the evolution of the universe filled
with a scalar field $\sigma$,  we must somehow set the initial
values of the field and its derivatives at different point in the
space. Also it is needed to specify the topology of the space and
its metric in a manner consistent with the initial values. In this
way, this kind of model is concerned with the evolution of a
universe filled with a chaotically distributed scalar field
$\sigma$.

In the context of chaotic inflation, another possible realization
is an axionic (shift)symmetry, whenever the inflaton is a
pseudo-scalar axion. In this scheme, the inflaton potential arises
due to the breaking of a (global) axionic symmetry, and it is
therefore controlled by it: for instance, the coupling of the
inflaton to matter does not affect the inflaton potential if the
axionic symmetry is respected. This mechanism, known as natural
inflation, was originally proposed in \cite{Free}, and several
possible implementations have been discussed in \cite{Adam}. Shift
symmetries also arise within string theory, and their application
to inflation has been considered for instance in \cite{Adam,Bank}.

Let us now consider  an effective potential given by
 $V=\lambda_n\,\sigma^n/n$ for $\sigma<\sigma_o$, which becomes
extremely steep at $\sigma>\sigma_o$. If the universe is created
at $\sigma$ just above $\sigma_o$, at the point
$V_o>V(\sigma_o)=\lambda_n\sigma_o^n/n$, the field immediately
falls down to $\sigma_o$ and acquires a velocity given by
$\dot{\sigma}_o^2/2=V_o-V(\sigma_o)$.

We will denote $V^\ast(\sigma_o)$ as the critical initial value of
$V$, such that inflation still exists for  the field $\sigma$
falling to the point $\sigma_o$ from the height $V_o<V^\ast$, and
 disappears at $V_o>V^\ast$. We also introduce the parameter $\beta_o$,
similar to the one of the previous section. During inflation, the
scale factor is given in the present situation by \be
\frac{a}{a_o}=\left(\frac{\phi}{\phi_o}\right)^{\gamma}=
\left(1+\frac{(\sigma_o^2-\sigma(t)^2)}
{\gamma\,n\,\varepsilon\phi_o^2}\right)^{\frac{\gamma}{2}},
 \en
and the universe expands
$\exp(\gamma/2\ln(1+\sigma_o^2/\gamma\,n\,\varepsilon\phi_o^2))$
times, when $\sigma(t)\longrightarrow$ 0.

We considered a model with $n=2$, where the effective potential
blows up at $\sigma>\sigma_o=0(10)$, $\omega=500$,
$8\pi\varepsilon\phi_o^2=1/2$,
$\dot{\widetilde{\phi_o}}^2=10^{-12}$ and $V=10^{-11}$. In this
case we have found that inflation begins when the $\sigma$ rolls
down to the point \be \sigma_{\inf}\approx\sigma_o+0.03+
0.14\ln\left[\frac{(1+6\varepsilon)\beta_o}{(1-6\varepsilon)}\right].
 \en
In order to achieve the 60 e-folds we first need  to give  the
values of $\varepsilon$ and $\phi_o$ so that we could  find an
initial value for the scalar field $\sigma$. For instance, if we
chose $\omega$=500 and $8\pi\varepsilon\phi_o^2=1/2$, we find that
$\sigma_o= 2.3$. With these values, and
 $V\sim10^{-11}$, the
probability of creation  of the universe is not suppressed for
$\beta_o\lesssim\,2.4\cdot10^{-10}$. Therefore one can argue that
it is most probable to have $\beta_o\sim\,2.4\cdot10^{-10}$. This
means that inflation typically starts in this model at
 \be
\sigma_{\inf}\,\approx\sigma_o-\,3.1, \en
 which means that inflation starts with a larger value  of the scalar field $\sigma$
 than the one obtained in the JBD theory.

Now we consider a model with $n=4$, where the effective potential
$V=\lambda_4\sigma^4/4$ blows up at $\sigma>\sigma_o=0(10)$,
$\omega=500$, $8\pi\varepsilon\phi_o^2=1/4$,
$\dot{\widetilde{\phi_o}}^2=10^{-12}$ and $V=10^{-11}$. In this
case we have found that inflation begins when the $\sigma$ rolls
down to the point \be \sigma_{\inf}\approx\sigma_o+0.07+
0.14\ln\left[\frac{(1+6\varepsilon)\beta_o}{(1-6\varepsilon)}\right].
 \en
As an example, we considered $\omega$=500 and
$8\pi\varepsilon\phi_o^2=1/4$, such that  $\sigma_o= 3.3$ is
required to obtain an universe that inflates $e^{60} $ times. Also
we have taken the values $V\sim10^{-11}$,
$\dot{\widetilde{\phi_o}}^2\sim10^{-12}$ and
$\beta_o\sim\,9.6\cdot10^{-10}$. This means that the inflation
field typically starts at
 \be
\sigma_{\inf}\,\approx\sigma_o-\,2.8. \en For $\sigma_o=$10,
inflation starts at $\sigma_{\inf}\sim7.2$, the
 universe inflates $e^{321}$ times and becomes flat. The universe
 inflates $e^{60}$ times for $\sigma_{\inf}=2.3$, i.e. for
 $\sigma_o=5.1$ and  this leads to $\Omega=1.1$.
 The numerical solution $\sigma(t)$ is shown in Fig.3 for some different
 values of the $\omega$ parameter and the same velocity,
 $\sigma_o$. In the same situation, we have studied the evolution
 of the JBD field $\phi$. We have found that this field
 monotonically increases towards some constant value, which is closer to
 the one determined by the actual value of the Planck mass
 (recall that $\varepsilon\phi_N^2=1/8\pi$), just when the
 inflaton scalar field $\sigma$ begins to oscillate near the
 minimum of the effective potential, located at $\sigma\approx$ 0.

\begin{figure}[ht]
\includegraphics[width=3.0in,angle=0,clip=true]{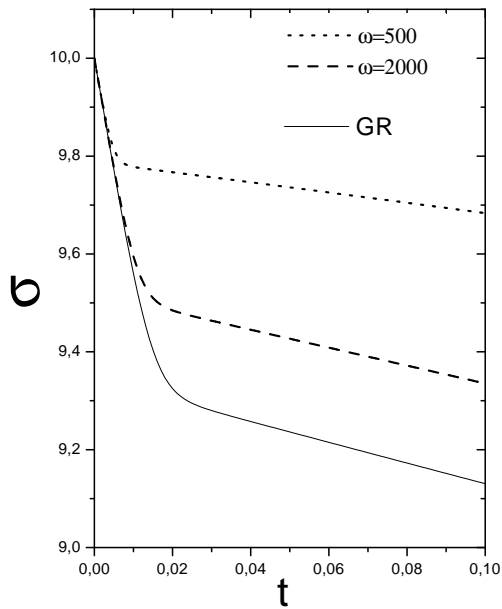}
\caption{This plot shows the inflaton field $\sigma$(t) as a
function of the cosmological times $t$, for the effective
potential $V(\sigma)=\lambda_4\,\sigma^4/4$, either in Einstein's
or in JBD theories ($\omega=500$ and $\omega=2000$) for the same
value of $\dot{\sigma_o}$. In these graphs we have assumed the
constant $\lambda_4$ to be 1. } \label{fig2}
\end{figure}

It is interesting to study our model density perturbations but, as
the JBD theory affects the density perturbations,  we calculated
the density perturbations in our model according to
Ref.~\cite{re11} and plotted the function

\be \ds \frac{\delta\,\rho}{\rho}\,\approx\,Cte\,H^{2}\left[
\left(\varepsilon\phi^{2}8\pi\right)^{\frac{3}{2}}\,\frac{1}{|\dot{\sigma}|}\,
+\frac{(1-\varepsilon\phi^{2}8\pi)}{2|\dot{\phi}|}\gamma\phi\right],
\label{ec10}
 \en
where
$$\ds \gamma\,=\,\frac{1}{\sqrt{\omega+3/2}},
$$
and $Cte=3/5\pi$, which corresponds to density perturbations in a
flat universe, and note that  equation (\ref{ec10}) coincides with
the GR, if we assume that $\varepsilon\phi^{2}=1/8\pi$. However,
these density perturbations should be supplemented by several
different contributions in a closed inflationary universe, which
may alter the final result for $\delta\rho/\rho$ at small N.
However we will  postpone a complete investigation of this problem
for the time being.

Fig.(4) shows the magnitude of the density perturbations as a
function of the  N e-folds parameter for the model $\lambda_4$,
for $\sigma_o=10$ and for various values of $\omega$. Similar to
the corresponding model in the Einstein's theory of GR ,
$\delta\rho/\rho$ has a maximum at small N $\simeq$ 0(7).
Numerically, we could show that it presents a small displacement
for $\omega$=500 with the maximum at N = 0(8) which corresponds to
a scale $\sim$ 10$^{25}$ cm, which is similar to that obtained in
the Einstein's GR theory.

\begin{figure}[ht]
\includegraphics[width=3.0in,angle=0,clip=true]{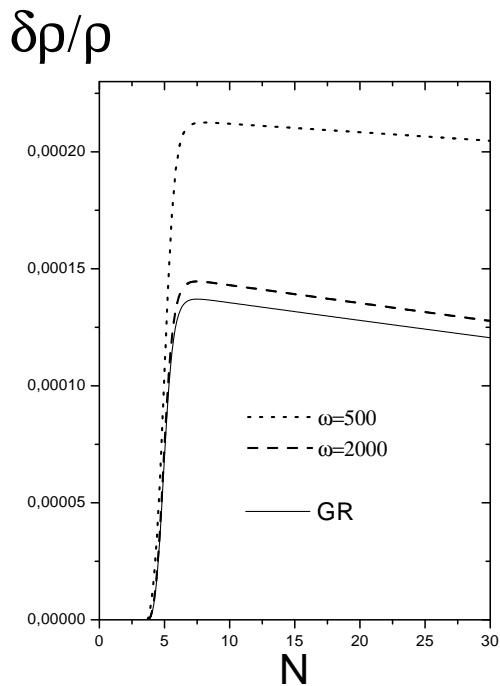}
\caption{Scalar density perturbation for the model
$V(\sigma)=\lambda_4\,\sigma^4/4$ as a function of the N e-folds
parameter for two different values of the JBD $\omega$ parameter.
The values that we have assumed are $\omega=500$ and $\omega=2000$
for $\lambda_4=10^{-14}$ and $\sigma_o=10$. These plots are
compared with that obtained by using Einstein's GR theory, where
$\delta\rho/\rho\approx\,Cte. H^2/|\dot{\sigma}|.$
       } \label{fig3}
\end{figure}

\section{conclusion}

 In this article we study a closed inflationary universe model
in a JBD theory. In Einstein´s GR theory this model was study by
Linde\cite{Lindeclosed}. Here, when $V$ is constant, in the
interval from $\sigma_o$ to $\sigma=0$ we found some problems.
First of all, and due to the form of the potential, it is found
that the universe typically either collapses very soon, or
inflates forever (graceful exit problem). Secondly, after the end
of inflation, which occurs when $\sigma$ reaches the value $\sigma
=0$, it seems that the inflaton field does not appear to engage
into oscillations. This is only because the model is
oversimplified. Actually, we should mention here that the
inspiration for this model comes from the Linde's Hybrid
inflationary model  where the end of inflation sends the system
towards a direction perpendicular to the inflaton direction in
field space (the direction of the waterfall). Then, both fields
begin oscillating around the true minimum of the scalar potential.
Thus, the problem mentioned above does not correspond to the
physical model, which is the basis for our toy-model.

In a JBD theory we have found that it is possible to successfully
solve the graceful exit problem for an effective potential $V=$
constant with an extremely steep at the Planck scale. The
dependence on the values of $\omega$ and the initial conditions
for $\phi_o$ and $\dot{\phi_o}$ permit us to reach the value
$\sigma=0$, needed for solving the graceful exit problem. These
conclusions are in good qualitative agreement with the result of a
numerical investigation.

We have found that for  some initial velocity values
$\dot{\sigma_o}$, it is probable  that in Einstein`s GR theory the
universe will have a large value of $\Omega$; however, in a JBD
theory and for $\omega$ parameter $\omega<2000$, it is possible to
obtain $\Omega\sim1.1$.

 Also,in the case in which the potential is very sharp at large
$\sigma$ we have found extra ingredients in the JBD theory, when
compared with its analog in Einstein's GR theory. Specifically, we
obtain a short stage of inflation for a closed universe. This
could be possible due to that the speed of the rolling field was
very large but it could be controlled with the value of the JBD
parameter $\omega$. Similarly, the long-wavelength perturbations
of the scalar field were suppressed. Both of these effects tend to
suppress the density perturbations produced at the beginning of
inflation, and they would be able to account for the suppression
of the large-angle CMB anisotropy observed by WMAP.

The $\delta\rho/\rho$ graphs present a small displacement with
respect to $N$, when compared with the results  obtained in
Einstein`s GR theory. This would change the constraint on the
value of the parameters  that appears in the scalar potentials
$V(\sigma)$. In this way, we have shown that closed inflationary
universe models in a JBD theory are less restricted than  one
analogous in Einstein`s GR theory   due to the introduction of a
new parameter, i.e. the $\omega$ parameter. The inclusion of this
parameter gives us a freedom that allows us to modify the
 cosmological model by simply modifying the corresponding value of
this parameter.

 As Linde mentioned in Ref.\cite{Lindeclosed}, the fine - tuning
required for the construction  of a closed inflationary model is
much  easier to achieve than the incredible fine - tuning required
for the explanation of the enormously large mass and entropy of
the universe,  its homogeneity and isotropy, and the observed
anisotropy of CMB without using inflation.

The models that we  have treated  are  fine-tuned, such that the
flatness of the universe remains   a  prediction of these
inflationary scenarios. Nevertheless one should keep in mind that
with the fine-tuning at a level of about one percent one can
obtain a semi-realistic model of an inflationary universe with
$\Omega>$ 1 Ref.\cite{Lindeclosed}.

\begin{acknowledgments}
SdC was supported from COMISION NACIONAL DE CIENCIAS Y TECNOLOGIA
through FONDECYT N$^0$s 1030469, 1040624 and 1051086 grants. Also,
it was partially supported by PUCV  by grant 123.764.
\end{acknowledgments}

\end{document}